\documentclass[11pt]{iopart}
\usepackage{bm}
\usepackage{hyperref}
\usepackage{amssymb}
\usepackage{amsopn}
\usepackage{slashed}
\RequirePackage{ifpdf}

\begin{document}

	\begin{flushright}
		\textsf{DESY 10-156}
	\end{flushright}
	\title{Atomic Precision Tests and Light Scalar Couplings}
	
	\author{Philippe Brax$^{1}$ and  Clare Burrage$^2$ }
	\vspace{3mm}
	
	\address{$^1$ Institut de Physique Th\'{e}orique,
	  CEA, IPhT, CNRS, URA2306, F-91191 Gif-sur-Yvette c\'{e}dex,
	  France \\[3mm]
	$^2$ 	  D\'{e}partment de Physique Th\'{e}orique,
	Universit\'{e} de Gen\`{e}ve, 24 Quai E. Ansermet,
	CH-1211, Gen\`{e}ve, Switzerland\\
	Theory Group,	 Deutsches Elektronen-Synchrotron DESY,	 D-22603,
	  Hamburg, Germany 	}

	\vspace{3mm}

	\eads{\mailto{philippe.brax@cea.fr},
	\mailto{clare.burrage@unige.ch}}
	
	\begin{abstract}
We calculate the shift in the atomic energy levels induced by the
presence of a scalar field which couples to matter and photons.  We
find that a combination of atomic measurements can be used to probe
both these couplings independently.  A new and stringent bound on the matter coupling springs from the precise measurement of the
1s to 2s energy level difference in the hydrogen atom, while the coupling to photons is essentially constrained by the Lamb shift. Combining these constraints with current particle physics bounds  we find  that the contribution of a scalar field  to the recently claimed discrepancy in the proton radius measured using electronic and muonic atoms is negligible.

	\end{abstract}
	\maketitle

	\section{Introduction}

	\label{sec:introduction}
Scalar fields coupled to matter occur in a wide variety of fundamental
contexts:  the inflationary scalar field should couple to matter in
order to reheat the universe, and such a coupling seems likely to  exist for
the dark energy scalar field too.  Indeed most attempts at modifying
gravity, or unifying it with particle physics, predict new scalars
coupling to the standard model particles in the form of
non-renormalisable interactions that are suppressed by the energy scale characteristic of
the energies probed by the model.  For example many coupled
scalars are present in the four
dimensional effective theories arising from compactifying the extra
dimensions of string theory \cite{Svrcek:2006yi,Arvanitaki:2009fg,Jaeckel:2010ni}.

It is well known that light canonical scalar fields which couple to
matter are tightly constrained by experimental searches for fifth
forces and violations of the equivalence principle \cite{Will:2001mx}, although
non-linear effects  such as in the chameleon \cite{Khoury:2003rn,Khoury:2003aq}, or Galileon \cite{Nicolis:2008in}
cases allow the fields to easily avoid these constraints
through dynamical mechanisms.   In this article we do not restrict ourselves to light
fields, specifying only that the field should not be so heavy that
its mass would lie above the cutoff of the low energy effective field
theory we wish to study.  These will be predominately atomic
experiments for which we study an effective field theory valid up to a
cut off high enough to include effects from the standard model of particle physics.  Therefore we consider the
effects of scalars with masses $\lesssim\mbox{ TeV}$ for which only
scalars with mass $\lesssim 10^{-5}\mbox{ eV}$ would violate
experimental bounds on the existence of fifth forces.

The phenomenology of such fields is not restricted to their effects in
gravitational experiments. The minimal scenario we consider is that the scalar field couples to matter
conformally, i.e.  through a scalar field dependent conformal
rescaling of the metric.  Classically a conformal coupling means that the scalar
field couples to fermions but not to massless bosons such as the
photon.  It was shown in \cite{Brax:2010uq, Brax:2009ey} however that, given a conformal
coupling,  quantum effects  lead to a
coupling between photons and scalars in the low energy effective
theory.  As already mentioned, in this article we focus our attention on
low energy atomic experiments which we consider to be described by low
effective theories with a cut off $\sim \mbox{ TeV}$.  We  include a coupling of
the scalar field to photons as a generic property of  the low energy theory.

  It is  also possible to study the effects
of coupled scalar fields both in particle colliders and in low energy, high precision experiments.  If
the scalar couples to photons it has a phenomenology similar to the
Peccei-Quinn axion, in particular oscillations between photons and
scalars can occur in the presence of  magnetic fields \cite{Raffelt:1987im}.  Much effort has gone into exploring the
consequences of such couplings both in the laboratory
\cite{PhysRevD.47.3707,Robilliard:2007bq,Chou:2007zzc,PhysRevD.78.092003,Fouche:2008jk,Afanasev:2008jt,Ehret2010149} and in astrophysics \cite{Raffelt:2006cw,DeAngelis:2007dy,Arik:2008mq,Burrage:2008ii,Burrage:2009mj,Fairbairn:2009zi,Mirizzi:2009aj,Bassan:2010ya}.
A scalar field coupled to the fields of the standard model  has also
direct effects on the properties of these particles - making the
masses of elementary particles, and other energy scales of the theory, become
dependent on the scalar field.  Consequently the
Hamiltonian describing the low energy behaviour of any fermion
becomes scalar field dependent.  Previously \cite{Brax:2010jk} we have shown that this
leads to  electrons transitioning between energy levels in atoms in
the presence of a  background scalar condensate. This gives  rise to new
possibilities of  searching for scalar fields  via
scalar field stimulated photon emission.  In this paper we show that the presence of a scalar
will also  perturb the
energy levels of atoms leading to new constraints on coupled scalar fields from precision atomic measurements.

The scalar field induced shift in energy levels can be constrained by measurements of the energy gap between the 1s and 2s orbitals of the
hydrogen atom.   A second consequence of the shifts in energy levels is a scalar field
dependent  change in the
Lamb shift:
the small difference in energy between the 2s and 2p energy levels of
hydrogen caused by the interaction between the electron and the
background \cite{PhysRev.72.241}.
We will show that the scalar field dependence of the Lamb shift,
induced by a coupling between fermions and the scalar field, will lead
to different values of the proton charge radius when measured with
muons and with electrons.  Comparison of properties of electronic and
muonic atoms will be shown to be a sensitive probe of the existence of
new scalar fields.
The recently  proposed existence of a possible
$5\sigma$ discrepancy between the proton charge radius\footnote{Effects of the finite size of the proton are relevant for s-states, because the wave function does not vanish at the location
of
 the proton, whereas p-states remain unaffected, and so the
 difference in energy
 between the 2s and 2p energy levels is sensitive to the dimensions of
 the proton. Muonic hydrogen  is more sensitive to this effect than electronic
 hydrogen because the heavier mass of the muon corresponds to a much
 smaller Bohr radius.}  measured
recently from the Lamb shift in muonic hydrogen (an atom formed by a proton and a
 negative muon), compared to that
inferred previously from hydrogen atom spectroscopy,  will lead to a new constraint on scalar couplings, i.e. an upper bound on the geometrical mean of the couplings to matter and photons.

Scalar fields are not the only new physics that can modify atomic spectra.  The presence of additional,
hidden sector, $U(1)$ gauge groups also give rise to a change in the
Lamb shift \cite{Pospelov:2008zw,Jaeckel:2010xx}, as well as
modifications of Coulomb's law which can be tested through atomic measurements.

In the following section, we recall standard properties of coupled scalars. In section \ref{sec:Atom}, we calculate the effect of a scalar field on the atomic levels, with  particular focus on
the energy level difference between the 1s and 2s states of the hydrogen atom and the Lamb shift. The 1s to 2s energy difference  leads to a new and stringent upper bound on the coupling of scalars to matter and the Lamb shift leads to a looser upper bound on the geometrical mean of the photon and matter couplings.
We then combine these atomic  constraints with those obtained from high energy particle physics experiments. We find that the upper bound on the matter coupling obtained from the 1s to 2s gap is stronger than the particle physics bounds. On the other hand the constraint on the coupling to photons deduced  from the width of the Z boson and electroweak precision tests  is more restrictive than atomic physics bounds. We then apply these results to  the proton charge radius discrepancy   and find that it is incompatible with the Z width bound on the coupling of scalars to gauge fields. Hence scalar fields cannot provide an explanation to the proton radius anomaly.  We conclude in section \ref{sec:Conc}.

\section{Coupled Scalar Fields}

We consider a scalar-tensor theory defined by the Lagrangian
\begin{equation}
S=\int d^4 x \sqrt{-g} \left(\frac{R}{2\kappa_4^2} -\frac{1}{2} (\partial
\phi)^2 -V(\phi)\right) + \mathcal{L}_m(\Psi_i, A^2(\phi)g_{\mu\nu}),
\label{eq:action}
\end{equation}
where the matter fields $\Psi_i$ feel a metric
$\tilde{g}_{\mu\nu}=A^2(\phi)g_{\mu\nu}$.    It is possible to allow the
scalar to couple differently to different particle species \cite{Brax:2009ey},
however we restrict ourselves here to a universal coupling which
captures all the relevant phenomenology.  Equation (\ref{eq:action})
is known as the Einstein action for a coupled scalar field, a
conformal rescaling allows a classically equivalent description of the
theory in which particle properties are independent of the scalar
field, but the gravitational sector of the theory becomes scalar
dependent, known as the Jordan frame description. For computational convenience we work in the Einstein frame in what follows.  We consider that the
Einstein frame theory (\ref{eq:action}) is an effective field theory
valid up to an energy cut off and that  couplings given by
non-renormalisable operators are  suppressed  by powers of the cut off scale.

In a non-relativistic background
the Klein-Gordon equation for the scalar field arising from (\ref{eq:action})  is modified and becomes
\begin{equation}
D^2\phi = \frac{\partial V}{\partial \phi} +     A(\phi)\rho ,
\label{eq:KG}
\end{equation}
where $\rho$ is the classical energy density.
The dynamics of the field $\phi$ are determined by the effective potential
\begin{equation}
V_{\rm eff}(\phi)= V(\phi) + \rho A(\phi).
\end{equation}
We assume that this effective potential has a minimum at $\phi=\phi_0$
so that the field is stabilised.

In addition to the terms in Equation (\ref{eq:action}) quantum effects
will generate \cite{Brax:2010uq,Brax:2009ey}  a term which describes the coupling of the scalar field to
photons
\begin{equation}
\mathcal{L}_{\gamma}= \frac{\phi}{2M_{\gamma}}F^{\mu\nu}F_{\mu\nu} =
\frac{\phi}{M_{\gamma}}(\mathbf{E}^2-\mathbf{B}^2),
\label{eq:Veff}
\end{equation}
where $F_{\mu\nu}$ is the electromagnetic field tensor, and
$\mathbf{E}$ and $\mathbf{B}$ are the electric and magnetic fields
respectively.  This contributes to the scalar field effective
potential in a similar manner to the background matter energy density
giving
\begin{equation}
V_{\rm eff}(\phi)= V(\phi) + \rho A(\phi) + \frac{\phi}{M_{\gamma}}(\mathbf{B}^2-\mathbf{E}^2).
\end{equation}
From this point onwards we assume that $\phi_0$ minimises the effective
potential including the background contributions from the electric and magnetic
fields.

In a homogeneous background  the scalar field  can be expanded about its minimum
value $\phi_0$
\begin{equation}
A(\phi)= A(\phi_0)\left(1 + \frac{A^{\prime}(\phi_0)}{A(\phi_0)}
  \delta \phi + \ldots \right),
\end{equation}
where $\delta\phi$ is the fluctuation induced by the presence of a
matter source.  We assume that higher order terms in this expression
are small and can be consistently neglected.  We can write the coefficient of the second term as
an inverse energy scale
\begin{equation}
\frac{A^{\prime} (\phi_0)}{A(\phi_0)}= \frac{1}{M_m}.
\end{equation}
We view (\ref{eq:action})  as a low energy energy effective theory,
and therefore expect  $M_m$ and $M_{\gamma}$  to be of the order of the cut off scale of
the theory in the relevant energy range as it will be sensitive to
fields that have been integrated out at higher energy.  We will be interested in an
effective low energy theory describing a  muon, or an electron  in the
background of a hydrogen nucleus, therefore  we expect $M_m,M_{\gamma} \gtrsim
\mbox{ GeV}$, larger than the muon and proton masses.
Particle physics effects at
accelerators and therefore the effective theory involving $W$ bosons  would require the coupling to $\phi$ to be determined by a
cut-off scale larger  than the mass of the W boson. Constraints from particle physics impose that
$M_m$ and $M_{\gamma}$  should be at least in the TeV  and $M_Z$ ranges respectively \cite{Brax:2009aw}.

It has been recently argued~\cite{Hui:2010dn} that the coupling of scalars to matter in scalar-tensor theories is preserved
under renormalisation and that the only effect of quantum corrections is to induce a change due to the wave function renormalisation of the scalar field. When integrating out momenta to a obtain the effective scalar theory valid at a lower energy scale,  radiative corrections  in the scalar sector imply that the wave function renormalisation $Z_\phi$ should be  affected by logarithmic terms and therefore
\begin{equation}
\phi_{\rm low} = Z_\phi \phi_{\rm high}
\end{equation}
where $\phi_{\rm low}$ is the normalised field after integration over momenta between the low energy and the high energy cut off. This leads to a direct relation between the coupling scale to matter at low and high energies
\begin{equation}
M_m ^{\rm low}= \frac{M_m^{\rm high}}{Z_\phi}
\end{equation}
Logarithmic corrections do not entail a large scale dependence of the
coupling scale $M_m$.

Within the effective field theory approach used in this
paper, the various coupling scales are just an effective
parametrisation which needs to be deduced from experiment.  In the absence of an
underlying theory going beyond the standard model coupled to scalars, we have simply given a useful parametrisation of the
scalar-matter coupling and its phenomenological consequences in atomic
physics.

The coupling of the scalar field to fermions implies that the fermion
masses $m_f$ become scalar field dependent:
\begin{equation}
m_f(\phi)= A(\phi) m_{f0},
\end{equation}
where $m_0$ is the bare mass as it appears in the Lagrangian.
Again expanding around the background value of the scalar field we find
\begin{equation}
m_f(\phi)= m_f \left(1+ \frac{ \delta \phi}{M_m}\right).
\end{equation}
We assume that the higher order terms in this expression can be
consistently neglected, and  we have  normalized $A(\phi_0)=1$. Notice that, viewed as a low energy operator the fermionic mass term
$m_f(\phi)\bar \psi\psi$ is a non renormalisable effective interaction term. When truncating this interaction to first order in $\phi$ it reduces to an effective Yukawa interaction with a coupling $m_f/M_m$ which must be small as the lepton of mass $m_f$ has not been integrated out.

\section{Atomic Energy Shifts}
\label{sec:Atom}
In and around atoms the scalar field perturbation is
sourced by the presence of the
nuclear electric field
\begin{equation}
{\bf E}= \frac{Ze{\bf r}}{4\pi r^3},
\end{equation}
implying a perturbation to the effective potential (\ref{eq:Veff})
\begin{equation}
\delta V= -\phi \frac{Z^2\alpha}{4\pi M_\gamma r^4},
\end{equation}
 and by the point-like density of the atomic  nucleus, $\delta \rho=
 m_N \delta^{(3)}$,  centrered at the origin and depending on the nuclear
mass $m_N$.
In spherical coordinates, the static scalar field perturbation then satisfies
\begin{equation}
\frac{d^2\delta \phi}{dr^2}+\frac{2}{r} \frac{d\delta\phi}{dr}= -\frac{  E^2}{M_{\gamma}}+ \frac{m_N}{M_m} \delta^{(3)},
\end{equation}
where we have neglected the scalar mass term. This approximation is valid as long as the range of the scalar force $1/m_\phi$ is larger than the size of the atom, implying that $m_\phi$ must be smaller than $10^4$ eV.
The solution which vanishes far away from the nucleus is
\begin{equation}
\delta\phi=- \frac{m_N}{4\pi M_m r}- \frac{Z^2\alpha }{8\pi M_{\gamma} r^2}.
\end{equation}
The sign of the scalar wave function is crucial as it leads to a
negative contribution to the energy levels. Notice that, except
extremely close to the nucleus, the first contribution dominates over
the second one.  For example, for a hydrogen atom the first term
dominates when $r>10^{-8} (M_m/M_{\gamma}) a_0$, with $a_0$ the Bohr
radius.  However  we retain the second term as we  will find that the
Lamb shift is independent of the scalar field perturbation due to the
mass of the nucleus.

At low energy, and in the non relativistic limit, the fermion wave function satisfies  a Schrodinger equation with the
interaction Hamiltonian given by \cite{Brax:2010jk}
\begin{equation}
H=\frac{p^2}{2m}+W +m-\frac{1}{2mM_m}\left(\delta\phi
  p^2+(\sigma\cdot p) \delta\phi (\sigma\cdot p)\right)  +\frac{m}{M_m}\delta\phi,
\label{Ham}
\end{equation}
where $m$ is the unperturbed fermion mass and    $W$ is a potential describing  the interactions  of the
fermion with all the other fields in the theory.
An order of magnitude estimate shows that the dominant  perturbation due to the scalar field  is
\begin{equation}
\delta H= \frac{m}{M_m}\delta\phi
\end{equation}
We consider the effect of the scalar field on the  $l=1$ and $l=2$  energy levels of a
hydrogenic atom with nuclear charge $Ze$.
The wave functions for these states are (in spherical polar coordinates)
\begin{eqnarray}
\psi_{1s}&=&\frac{1}{\sqrt{\pi}}\left(\frac{Z}{a_0}\right)^{3/2}e^{-Zr/a_0},\\
\psi_{2p}&=&\frac{1}{\sqrt{\pi}}\left(\frac{Z}{2a_0}\right)^{5/2}e^{-Zr/2a_0}r\cos\theta,\\
\psi_{2s}&=& \frac{1}{4\sqrt{2\pi}}\left(\frac{Z}{a_0}\right)^{3/2}\left(2-\frac{Zr}{a_0}\right) e^{-Zr/2a_0},
\end{eqnarray}
where $a_0=\hbar/m_f c\alpha$ is the Bohr radius, with  $m_f$
the mass of the fermion in the orbital, $c$  the speed of light, and
$\alpha$ the fine structure constant.
The corresponding energy levels are shifted by the effects of the
scalar field,
\begin{eqnarray}
\delta E_{1s}&=&-\frac{Zm_N }{4\pi M_m^2 a_0} m-\frac{Z^4\alpha }{4\pi a_0^2M_mM_{\gamma}}m,\label{eq:E1}\\
\delta E_{2s}&=& -\frac{Zm_N }{16\pi M_m^2 a_0} m-\frac{Z^4\alpha }{32\pi a_0^2M_mM_{\gamma}}m,\\
\delta E_{2p}&=& -\frac{Zm_N }{16\pi M_m^2 a_0} m-\frac{Z^4\alpha }{96\pi a_0^2M_mM_{\gamma}}m.\label{eq:E3}
\end{eqnarray}
Notice that the higher levels are less affected by the scalar field but that  the gap
between the levels has increased. We will study two main effects:  the energy gap between the 1s and 2s levels, and the
Lamb shift which involves the difference of energy between the 2s and
2p levels. It is
clear from  Equations
(\ref{eq:E1})-(\ref{eq:E3}) that the Lamb shift is  sensitive only  to
the electric field contribution to the scalar wave function, and
that the gap between the 1s and 2s levels is essentially due to the nuclear point mass.

\subsection{Precision Measurements of Hydrogenic Atoms}

A strong constraint on $M_m$ can be deduced using the precision
measurements of hydrogenic energy levels, as  a low value
of $M_m$ would lead to large observable shifts.
The 1s-2s
transition for a standard hydrogen atom has a total uncertainty (experimental and theoretical) of
order $10^{-9}$ eV at the 1-$\sigma$ level~\cite{Schwob:1999zz,Simon:1980hu,Jaeckel:2010xx}. This transition receives a
contribution from the scalar field
\begin{equation}
\delta E_{1s-2s}= \frac{3 m_N }{16\pi M_m^2 a_0} m_e + \frac{7\alpha }{32\pi a_0^2M_{}^2}m_e,
\end{equation}
where we have defined $M^2=M_m M_\gamma$.
The contribution of the term sourced by the atomic electric field satisfies the bound if $\mbox{GeV} \lesssim M$.  However, the nuclear mass term  exceeds the $10^{-9}$ eV bound unless
\begin{equation}
M_m\gtrsim 10~ {\rm TeV}.
\label{eq:Mbound}
\end{equation}

\subsection{The Lamb Shift and the Proton Radius}

A second crucial effect of the coupling of the
fermions to the scalar field is the change in the
energy difference between levels with differing  angular momentum
$l$. The most important case is $l=2$, and the contribution to the Lamb shift is
\begin{equation}
\delta E_{2s-2p}= \frac{Z^4\alpha }{48\pi a_0^2M_{}^2}m.
\end{equation}

The change in the Lamb shift induced by the scalar field will vary
between electronic and muonic atoms.  For $Z=1$, we find that for an
electron with mass $m_e=0.51\mbox{ MeV}$ and Bohr radius $5.3\times
10^{-11} \mbox{ m}$ the scalar contribution to the Lamb shift is
\begin{equation}
\delta E_{2s-2p}(e^-)=3\times10^{-10} \left(\frac{\rm{GeV}}{M_{}}\right)^2
\mbox{ eV}.
\end{equation}
For a muon of mass $m_{\mu}=106\mbox{ MeV}$ and Bohr radius $2.5
\times 10^{-13}\mbox{ m}$ the contribution to the Lamb shift is
\begin{equation}
\delta E_{2s-2p}(\mu^-)=3\times10^{-3} \left(\frac{\rm{GeV}}{M_{}}\right)^2
\mbox{ eV}.
\end{equation}

As discussed in the introduction the Lamb shift can be used to infer
the charge radius of the proton measured in femtometers \cite{Pohl:2010zz}
\begin{equation}
\frac{\Delta E}{\mbox{meV}} = 210 - 5.23
  \left(\frac{r_p}{\mbox{fm}}\right)^2+0.035\left(\frac{r_p}{\mbox{fm}}\right)^3.
\end{equation}
Measurements of the Lamb shift may give different values for the
charge radius depending on whether the experiments are conducted with
electronic or muonic atoms. The CODATA value $r_p=0.8768\pm 0.0069 \mbox{ fm}$
\cite{Mohr:2008fa} is extracted mainly from spectroscopy of electronic hydrogen
atoms and is in agreement with the calculations of bound state quantum
electrodynamics \cite{Eides:2000xc,Karshenboim:2005iy}.  Ensuring that
the electronic Lamb shift lies within the current experimental limits
requires  $ 10^{-4}\mbox{ GeV}\lesssim M_{}$.  Then
the muonic Lamb shift  corresponds to a negative variation of the proton radius,
\begin{equation}
\frac{\delta r_p(\mu^-)}{r_p}=-0.4 \left(\frac{{\rm GeV}}{M_{}}\right)^2,
\end{equation}
So that the proton charge radius could vary between measurements with muons and with electrons.

The recent measurement of the proton charge radius for muonic hydrogen \cite{Pohl:2010zz}
gives $r_p=0.84184$ fm
although we
note  that it may yet be possible to explain this seemingly
anomalous measurement  with conventional
QED \cite{DeRujula:2010dp} and QCD \cite{Friedmann:2009mz}.  This corresponds to a negative variation of order four
percent, and  would require a suppression scale
\begin{equation}
M_{}\approx 3.2~ {\rm GeV}.
\label{eq:M}
\end{equation}
This is a reasonable scale for an effective theory at low energy
which includes protons in its spectrum. Of course, larger values of $M$ lead to a smaller contribution to the proton radius. The bound on $M_m$ obtained in (\ref{eq:Mbound}) is much larger than the value of the averaged scale $M$ deduced
from the proton radius deviation, implying $M_{\gamma}\lesssim
10^{-3}\mbox{ GeV}$.  In the following section we analyse whether such values are compatible with high energy particle physics experiments.

The Lamb
shift for hydrogenic atoms can also be used to constrain $M$. For
$Z=2$, the 2-$\sigma$ theoretical and experimental uncertainty is
$3\times10^{-9}$ eV \cite{vanWijngaarden:1991zz}, while for $Z=15$
it is $6\times10^{-4}$ eV~\cite{0295-5075-5-6-005,Johnson:2001nk} at the one-$\sigma$ level and the theoretical uncertainty is 8 eV for $Z=110$ \cite{Johnson:2001nk}. For $M=3.2$ GeV, we find that
the scalar contributions are respectively $5\times10^{-10}$ eV,
$2\times 10^{-6}$ eV and $4\times 10^{-4}$ which are within these bounds.
Hence we find that the constraint on $M$ coming from the proton radius of muonic atoms is compatible with high precision atomic tests for hydrogenic atoms. Larger values of $M$ would lead to even smaller contributions from the scalar field.

We have obtained a strong constraint on $M_m$ from the 1s-2s energy
gap of the hydrogen atom, which implies a constraint on $M_{\gamma}$ through measurements of the charge radius of muonic hydrogen. Independent
constraints on $M_\gamma$ can be deduced from optical cavity
experiments and astrophysical observations which probe near vacuum
environments.  Optical cavity experiments constrain scalar fields
with $m_{\phi}\lesssim \mbox{ meV}$ to have $M_{\gamma}\gtrsim 10^7 \mbox{ GeV}$
\cite{PhysRevD.47.3707,Chou:2007zzc,Ehret2010149}\footnote{ These results have been recently extended to a larger mass range up to $0.1$ eV by the GammeV-CHASE group. For masses larger than $10^{-3}$ eV, the bound on $M_\gamma$ becomes looser and looser up to a scale of order $0.1$ eV.}.   Astrophysics  constrains
$M_\gamma\gtrsim 10^9 \mbox{ GeV}$  for masses of the scalar less than
$10^{-12}$ eV in the interstellar medium
\cite{Burrage:2008ii}. Stronger constraints from helioscope
experiments and the alteration of the star burning rate would apply if
scalars were produced in the very dense environment inside stars.
However,
the dependence of the properties of the scalar field on the density of
its environment, as in  (\ref{eq:KG}),
implies that scalars are difficult to produce inside stars.  This was
first noticed in \cite{Brax:2007ak,Jaeckel:2006xm} in the chameleon context. In fact
as shown in \cite{Brax:2010xq} the production of scalars inside a
dense plasma can only be realised when the mass of the scalar is tuned
to be resonant with the plasma frequency. In other cases, scalars are
most likely to be very difficult to produce in stellar plasmas.  For
this reason, we only consider here constraints on $M_{\gamma}$ coming
from near vacuum experiments, and therefore for scalar fields with masses in
vacuum $m_{\phi}\gtrsim 10^{-3} \mbox{ eV}$  a coupling $M_\gamma< 10^{-3} \mbox{ GeV}$ as required to explain the charge radius of muonic hydrogen is permitted by optical experiments.

\subsection{Anomalous Magnetic Moment, the Z width and electro-weak precision tests}
\label{sec:highenergy}

So far, we have only considered atomic precision tests. In this subsection, we will confront the bounds deduced from atomic physics with the ones obtained in particle physics.
A very stringent  particle physics precision test which could be affected by the coupling of scalar fields to fermions is the anomalous magnetic moment
$g_f-2$ \cite{Jegerlehner:2007xe}. Typically, the measurement of the anomalous moment of fermions such as the muon involves the decay of pions via weak interactions with
resulting ${\cal O}(1)~ {\rm GeV}$ scale muons.   Therefore the cut
off of the effective theory describing this decay must lie above the
weak scale.  For such experiments, the contributions of scalars can be calculated
\begin{equation}
g_f-2 =\frac{1}{(4\pi)^2}\frac{m_f^2}{ M_{m}^2}\ln \frac{ M_{m}^2}{m_f^2}
\end{equation}
A suppression of $ M_{m}\approx 600$ GeV would lead to
$g_\mu-2 \approx 3\times  10^{-9}$ and would explain  the discrepancy
between the standard model prediction  and the
measured value of the muonic anomalous magnetic moment
\cite{Jegerlehner:2007xe,Hagiwara:2006jt}. In practise the constraint
$M_m\gtrsim 10$ TeV obtained from the 1s-2s gap of hydrogenic atoms
is a stronger constraint on the strength of the coupling than that from $g_f -2$ of the muon.

Gauge invariance at high energy implies that after electroweak symmetry breaking, the coupling scale to photons and the Z boson are the same at the weak scale. Radiative corrections between the weak scale and the QCD scale will not lead to large effects and the two couplings will be essentially the same. Now the coupling of the Z boson to a light scalar implies that the width of the Z boson is affected. Known bounds on the Z width can be applied to the branching ratio \cite{Brax:2009aw}
\begin{equation}
\frac{\Gamma( Z\to \phi f \bar f)}{\Gamma (Z\to f\bar f)}\approx \frac{1}{80\pi^3} \frac{m_Z^2}{M_\gamma^2}
\end{equation}
where $f\bar f$ is a fermion pair, and $m_Z$ is the mass of the Z boson. The uncertainty on the width is of order $0.0023$ GeV compared to a central value of $2.4952$ GeV. This leads to a bound $M_\gamma \gtrsim 60 $ GeV \cite{Brax:2009aw}. A stronger bound $M_\gamma\gtrsim \mbox{ TeV}$ can also be deduced from the electro-weak precison tests \cite{Brax:2009aw}. This implies the bound $M\gtrsim 3$ TeV  meaning that a scalar coupling to both matter and photons cannot explain the observed discrepancy in the proton radius. At most we find that the effect of scalar on the proton radius must be
\begin{equation}
\vert \frac{\delta r_p (\mu^-)}{r_p}\vert \le 10^{-6}
\end{equation}
Of course, this is a completely unobservable result.

\section{Conclusions}
\label{sec:Conc}

We have shown that scalar fields coupled to matter will shift atomic energy levels.  Combinations of atomic precision measurements of electric and muonic hydrogen can be used to probe both the coupling of the scalar field to matter and to photons.
The shifts follow from the form of the coupling of scalars to
fermions and photons, i.e. a non-renormalisable operator  with a
suppression scale which strongly depends on the cut off energy  below
which  the effective Lagrangian description is valid. It also depends
on the two types of sources for the scalar field perturbation inside
an atom: the nuclear energy density and the nuclear electric field.
We find that only  the contribution from the electric field has  a
direct effect on the Lamb shift, but that the nuclear energy density
has an effect on the 1s-2s energy gap of hydrogenic atoms.

Particle physics measurements  at accelerators have previously been shown to  require  a large suppression scale for the
coupling of the scalar field to the gauge sector of the standard model. We have shown
that the constraint deduced from measurements of the 1s-2s energy gap of the hydrogen
atom is  stronger and constrains $M_m\gtrsim 10$ TeV. We have obtained  that the scalar field perturbation of the Lamb shift is sensitive only to the product of the coupling scales $M_m M_{\gamma}$, and that the scalar field perturbation is much larger for muonic that electric hydrogen atoms.  However constraints on $M_{\gamma}$ coming from the electro-weak precision tests,  $M_\gamma \gtrsim 1$ TeV,  imply that the anomalous  measurement of the muonic
Lamb shift cannot be explained through the presence of a scalar field. This would have required   a low value of
the coupling scale of the scalar to photons $M_{\gamma}\sim \mbox{
  MeV}$.

In summary, we have shown  that the best atomic and particle physics bounds on the coupling scales of matter and photons to a scalar field are $M_m\gtrsim 10$ TeV and $M_\gamma\gtrsim 1$ TeV. Although we have found that the Lamb shift in hydrogenic and muonic atoms differ due to the difference of mass between the electron and the muon, the large values of $M_m$ and $M_\gamma$ imply that the shift in the proton radius due to a scalar field is negligible. Therefore it seems that the best possibility of detecting a coupled scalar field in laboratory experiments is not with atomic measurements but with  optical cavity experiments.

\section*{Acknowledgments}
We would like to thank Andreas Ringwald for very helpful
discussions in the preparation of this work.  CB was supported by the German Science Foundation (DFG) under the
Collaborative Research Centre (SFB) 676 and by the SNF.

\section*{References}
	\bibliographystyle{JHEP}
\bibliography{muon_final}

\end{document}